\documentclass[prb,aps,superscriptaddress,twocolumn,driverfallback=dvips]{revtex4}
\usepackage{feynmp}
\usepackage{amsmath}
\usepackage{amssymb}
\usepackage{epsfig}
\usepackage{graphicx}
\usepackage{subfigure}
\usepackage{hyperref}
\usepackage{bbm}
\usepackage{epsfig}
\usepackage{color}
\usepackage{float}
\usepackage{mhchem}
\usepackage{epstopdf}
\usepackage{caption}
\usepackage{booktabs}

\newcommand{\Rmnum}[1]{\expandafter\@slowromancap\romannumeral #1@}
\allowdisplaybreaks[3]

\def\f{\frac}

\def\G{\Gamma}

\def\m{\mathbf}

\def\w{\omega}

\def\e{\epsilon}

\def\mr{\mathrm}
\def\n{\nonumber}

\begin{document}
\begin{sloppypar}
\title{Superconductivity near an Ising nematic quantum critical point in two dimensions}

\author{Jie Huang}
\affiliation{Department of Modern Physics, University of Science and Technology of China, Hefei, Anhui 230026, China}
\author{Zhao-Kun Yang}
\affiliation{Department of Modern Physics, University of Science and Technology of China, Hefei, Anhui 230026, China}
\author{Jing-Rong Wang}
\affiliation{High Magnetic Field Laboratory of Anhui Province, Chinese Academy of Sciences, Hefei 230031, China}
\author{Guo-Zhu Liu}
\altaffiliation{Corresponding author: gzliu@ustc.edu.cn}
\affiliation{Department of Modern Physics, University of Science and Technology of China, Hefei, Anhui 230026, China}

\begin{abstract}
Near a two-dimensional Ising-type nematic quantum critical point, the quantum fluctuations of the nematic order parameter are coupled to the electrons, leading to non-Fermi liquid behavior and unconventional superconductivity. The interplay between these two effects has been extensively studied through the Eliashberg equations for the superconducting gap. However, previous studies often rely on various approximations that may introduce uncertainties in the results. Here, we re-visit the issue of how the superconducting transition temperature $T_{c}$ is affected by removing certain common approximations. We numerically solve the self-consistent Dyson-Schwinger equations of the electron propagator $G(p)$, the nematic propagator $D(q)$, and the vertex function $\Gamma_{\mathrm{v}}^{\mathrm{1L}}(p+q,p)$ expanded up to the triangle order, without introducing further approximations. Our calculations reveal that the extended $s$-wave superconducting gap is the only convergent solution to the nonlinear gap equations. We investigate the evolution of $T_{c}$ as the system approaches the nematic quantum critical point from the disordered (tetragonal) phase. Under the bare vertex approximation, $T_{c}$ is monotonically enhanced. However, when vertex corrections are incorporated, $T_{c}$ initially increases but then decreases, with the maximum value of $T_{c}$ occurring at a point away from the quantum critical point. The obtained gap symmetry and the non-monotonic behavior of $T_{c}$ are compared with recent experiments on doped FeSe materials.
\end{abstract}

\maketitle


\section{Introduction \label{Sec:introduction}}

The electronic nematic order refers to a state of quantum matter that spontaneously breaks the $C_{4}$-symmetry down to a $C_{2}$-symmetry. Investigations of this order date back to Kivelson, Fradkin, and Emery \cite{Kivelson98}, who predicted the existence of electronic nematicity in normal states of cuprate superconductors as a result of the stripe melting. Formation of electronic nematic order can also be studied from a different perspective \cite{Metzner03, Oganesyan01}, namely the Pomeranchuk instability of the Fermi surface. When a system exhibits nematic order, its resistivity and other observable quantities become strongly anisotropic. Experimentally, clear signatures of electronic nematic order have been observed in many condensed matter systems, including cuprates \cite{Ando02, Fradkin15, Keimer15, Hinkov08, Daou10, Lawler10, Sato17, WuJ17, Gallais, Davis, Ishida20}, iron-based superconductors \cite{Chubukov12, Fernandes14, Shibauchi14, Coldea17, Bohmer17, Chuang10, ChuJiunHaw10, Song11, Kasahara12, LuXingYe14, Baek14, Kuo16, Hosoi16, Sunjp16, Matsuura17, Wiecki18, Sato18, Hanaguri18, Coldea20, Worasaran21, Mukasa21, Gallais21, Shibauchi22, Occhialini23, Mukasa23, Matsuura23, Coldea24, Dingqp25, Shibauchi25}, some heavy-fermion superconductor \cite{Ronning17}, Sr$_{2}$RuO$_{4}$ \cite{WuJie2020}, Kagome superconductors \cite{NieLinPeng22, WuPing23, HuYong23}, magic-angle graphene \cite{CaoYuan21, Jiang19}, superconducting nickelate \cite{JiHaoRan23}, and Ba$_{1-x}$Sr$_{x}$Ni$_{2}$As$_{2}$ \cite{Eckberg19}. Many of these superconductors share two common features: the presence of a dome shape for the superconducting transition temperature $T_{c}$ and the emergence of non-Fermi liquid (NFL) behavior in the normal state.

In iron-based superconductors, the electronic nematic order almost always coexists with antiferromagnetism or charge-density-wave order, making it hard to disentangle the intrinsic effect of the electronic nematicity on superconductivity. FeSe \cite{Coldea17, Bohmer17, Hosoi16, Sunjp16, Matsuura17, Sato18, Hanaguri18, Wiecki18, Coldea20, Gallais21, Mukasa21, Shibauchi22, Occhialini23, Mukasa23, Matsuura23, Coldea24, Dingqp25, Shibauchi25} has attracted special interest because this material displays nematic order but is not magnetically ordered. At ambient pressure, FeSe undergoes a nematic transition at the critical temperature $T_{s}\approx 90$ K without any long-range magnetic order. Chemical substitution of Se by S or Te tunes $T_{s}$ down to zero, yielding a well-defined nematic quantum critical point (QCP) at $x_{c} = 0.17$ in FeSe$_{1-x}$S$_{x}$ and at $x_{c}=0.50$ in FeSe$_{1-x}$Te$_{x}$. Nematic QCP can also be realized when the pressure \cite{Matsuura17, Coldea20, Coldea24} applied to FeSe$_{1-x}$S$_{x}$ exceeds certain threshold $p_{c}$. These nematic QCPs provide an ideal platform for exploring the exclusive impact of the nematic quantum criticality on both superconductivity and NFL behavior.

Now consider the nematic QCP defined at a critical non-thermal parameter $x=x_{c}$. Without lose of generality, the system is supposed to be tetragonal for $x>x_c$ and orthohombic for $x<x_{c}$. While the nematic order parameter has a vanishing vacuum expectation value ($\langle\phi\rangle = 0$) at the QCP, nematic quantum fluctuations are significant and interact strongly with gapless electrons near the Fermi surface. This interaction can be modeled by a Yukawa-type fermion-boson coupling. The physical effects of this coupling have been studied for decades by employing various analytical and numerical techniques \cite{Oganesyan01, Metzner03, DellAnna06, Rech06, Metlitski2010, Lederer15, Kivelson16, Kivelson17, Metlitski15, Scalapino15, Einenkel15, Raghu15, Chubukov16, Yamase13, Yamase16, Vishwanath16, Wanghj18, Klein18, Klein19, Wuyiming19, Damia19, Klein20, Damia20, Damia21}. Many of these calculations predicted that this coupling induces NFL behavior at high temperatures and unconventional superconductivity at low temperatures under appropriate conditions. Especially, the superconductivity mediated by nematic fluctuations was found to be significantly enhanced at the nematic QCP and suppressed as the system is tuned away from the QCP \cite{Lederer15, Kivelson16, Kivelson17, Metlitski15, Scalapino15, Yamase13}, implying that optimal superconductivity is realized right at the nematic QCP. These results are anticipated to offer a promising explanation for the two aforementioned common features of unconventional superconductors.

However, recent experiments have provided only limited support for the theoretical predictions mentioned above. The evolution of $T_{c}$ across the nematic QCPs is not universal, but material dependent. In FeSe$_{1-x}$S$_{x}$, $T_{c}$ is hardly modified as the system is tuned across the QCP. For FeSe$_{1-x}$Te$_{x}$, although a superconducting dome was observed near the nematic QCP \cite{Mukasa21, Shibauchi22, Mukasa23}, the value of $T_{c}$ is not maximized exactly at the QCP. There is not a well-established explanation for such an anomalous dependence of $T_{c}$ on the distance to nematic QCP.

In this paper, we revisit the quantum criticality of an Ising-type nematic transition in two spatial dimensions and study the properties of the superconductivity that emerges from the NFL state near the nematic QCP. The nematic order is assumed to have a $d$-wave symmetry, characterized by an anisotropic form factor in the Yukawa coupling term \cite{Lederer15, Klein18}. For simplicity, we only consider the disordered side of nematic transition, where $x \geq x_c$. In the limits of $x \gg x_{c}$ and $x\rightarrow x_c$, the system behaves as a normal FL and a typical NFL, respectively. Within the broad intermediate range of $x$, the system resides in a mixed FL/NFL regime, exhibiting FL behavior at low energies and NFL behavior at high energies.

NFL behavior and superconductivity are both induced by the same Yukawa coupling, implying that they can mutually influence each other. Consequently, they should be treated on an equal footing. In some previous theoretical works \cite{Lederer15, Chubukov16, Wanghj18, Klein18, Klein19, Wuyiming19, Damia20, Damia21}, their interplay has been studied using the Migdal-Eliashberg (ME) theory \cite{Migdal, Eliashberg}, which neglects all the vertex corrections to the Yukawa coupling. From a technical perspective, solving the self-consistent ME equations for renormalization factors and pairing function is challenging because they are highly nonlinear and have several variables. To reduce computational difficulty, a series of approximations have been introduced in practical computations. For instance, the nonlinear equations are often linearized in the vicinity of $T_{c}$. The boson self-energy, also called polarization function, is typically computed perturbatively under random phase approximation (RPA). Moreover, the electron momentum is usually restricted to the Fermi surface. While these approximations substantially simplify the original nonlinear equations, they may discard subtle feedback between NFL behavior and Cooper pairing and therefore lead to uncertainties in the predicted value of $T_{c}$.

Here, we study the superconducting pairing mediated by nematic quantum fluctuations by incorporating the vertex corrections into the ME theory. To describe NFL behavior and Cooper pairing equally, we employ the self-consistent Dyson-Schwinger (DS) equations satisfied by the renormalized electron propagator $G(p)$, the renormalized nematic propagator $D(q)$, and the vertex function $\Gamma^{\mathrm{1L}}(p+q,p)$ that includes the contribution of the lowest order vertex corrections. We convert these DS equations to four nonlinear integral equations for the wave-function renormalization $A_{1}(p)$, the mass renormalization factor $A_{2}(p)$, the pairing function $\Phi(p)$, and the boson polarization $\Pi(q)$. These equations fully capture the mutual influence of various quantum critical phenomena, such as Landau damping, electron mass renormalization, and superconducting pairing. We then solve these equations numerically in a self-consistent manner, without adopting linearization and Fermi-surface approximation. Our numerical calculations show that extended $s$-wave gap is the only stable convergent solution of the nonlinear equations. Strikingly, we find that the inclusion of vertex corrections can qualitatively change the behavior of $T_{c}$. Within the bare-vertex approximation, $T_{c}$ is enhanced monotonously as the QCP is approached. After including vertex corrections, $T_{c}$ exhibits a non-monotonic dependence on the boson mass parameter $r$ that measures the distance of the system to nematic QCP where $r=0$: $T_{c}$ first increases and then decreases as $r$ is lowering. We demonstrate that the non-monotonicity of $T_{c}$ originates from a complicated interplay between NFL behavior and Cooper pairing. This theoretical result provides a possible explanation of the non-monotonic doping dependence of $T_{c}$ observed in the tetragonal phase of FeSe$_{1-x}$Te$_{x}$ superconductor \cite{Mukasa21, Shibauchi22, Mukasa23}.

The rest of the paper is arranged as follows. In Sec.~\ref{sec:model}, we define the effective model of nematic quantum critical systems and obtain the self-consistent DS integral equations for $A_{1}(p)$, $A_{2}(p)$, $\Phi(p)$, and $\Pi(q)$. In Sec.~\ref{sec:gapsymmetry}, we present the numerical solutions of these equations and then analyze their physical implications. The theoretical results are compared with experiments on FeSe family superconductors. In Sec.~\ref{sec:summary}, we briefly summarize the results and highlight further projects.

\section{Effective model \label{sec:model}}

We consider a $(2+1)$-dimensional field theory that describes the interaction between the gapless electrons and the nematic quantum fluctuations, which is expressed as
\begin{eqnarray}
\mathcal{L} &=& \psi^{\dagger}\big(i\epsilon_{n} -
\xi_{\mathbf{p}}\big)\psi - g\hat{f}(\mathbf{p})\phi
\psi^{\dagger}\psi \nonumber \\
&& + \phi^{\dag}\big((i\omega_{m})^{2}/v_{\mathrm{B}}^2 -
\mathbf{q}^{2}-r\big)\phi, \label{eq:effectivemodel}
\end{eqnarray}
where $\psi$ and $\phi$ are the field operators of electrons and nematic fluctuations, respectively. Within the Matsubara formalism, for notational simplicity, the electron and boson energy-momenta are denoted by $p = (i\epsilon_n,\mathbf{p})$ and $q = (i\omega_m, \mathbf{q})$, respectively. The electron frequencies are $\epsilon_{n}=(2n+1)\pi T$ and the boson frequencies are $\omega_{m}=2m\pi T$.  The electron dispersion is $\xi_{\mathbf{p}}=\frac{\mathbf{p}^{2}}{2m}-E_{\mathrm{F}}$, where $E_{\mathrm{F}}$ is the Fermi energy. The boson mass $r$ serves as the non-thermal parameter that tunes nematic quantum phase transition at zero temperature, with $r=0$ marking the nematic QCP. Our current interest is restricted to the disordered side of nematic transition, corresponding to positive values of $r$. The precise value of boson velocity $v_{\mathrm{B}}$ is currently unknown. It is generally assumed \cite{Zhangss24} that $v_{\mathrm{B}}$ is of the same order of magnitude as the electron Fermi velocity $v_{\mathrm{F}}$. For simplicity, we set $v_{\mathrm{B}} = v_{\mathrm{F}}$.

The boson self-coupling term $(\phi^{\dagger}\phi)^{2}$ is usually an irrelevant perturbation, especially near the QCP, and thus is not included. The effective strength of the Yukawa coupling is strongly anisotropic, characterized by a constant $g$ and an angle-dependent form factor $\hat{f}(\mathbf{p})$. Following the works of Lederer \emph{et al.} \cite{Lederer15} and Klein \emph{et al.} \cite{Klein18, Klein19, Klein20}, we consider the following $d$-wave form factor:
\begin{eqnarray}
{f}(\mathbf{p}) = \frac{p_{x}^{2}-p_{y}^{2}}{p_{x}^{2}+p_{y}^{2}}.
\end{eqnarray}

Using the standard Nambu spinor \cite{Nambu}, denoted by $\Psi$, and the Pauli
matrices, we re-write the Lagrangian density as
\begin{eqnarray}
\mathcal{L} &=& \Psi^{\dagger}\big(i\epsilon_{n}\sigma_{0} -
\xi_{\mathbf{p}}\sigma_{3}\big)\Psi - g{f}(\mathbf{p})\phi
\Psi^{\dagger}\sigma_{0}\Psi \nonumber \\
&& + \phi^{\dag}\big((i\omega_{m})^{2}/v_{\mathrm{B}}^2 -\mathbf{q}^{2} - r\big)\phi. \label{nematiclagrangiannambu}
\end{eqnarray}
The free electron propagator is
\begin{eqnarray}
G_{0}(p) \equiv G_{0}(\epsilon_{n},\mathbf{p}) =
\frac{1}{i\epsilon_{n}\sigma_{0} - \xi_{\mathbf{p}}\sigma_{3}},
\end{eqnarray}
and the free boson propagator is
\begin{eqnarray}
D_{0}(q) \equiv D_{0}(\omega_{m},\mathbf{q}) =
\frac{1}{(i\omega_{m})^{2}/v_{\mathrm{B}}^2 -{\m q}^{2}-r}.
\end{eqnarray}
The free and fully renormalized electron propagators satisfy the following DS equation
\begin{eqnarray}
G^{-1}(p) &=& G_{0}^{-1}(p) + g^{2}\int_{q} {f}^{2}
\left(\m{p}+\f{\m{q}}{2}\right)D(q) \nonumber \\
&&\times \sigma_{3}G(p+q)\Gamma_{\mathrm{v}}(p+q,p),
\label{eq:dseofgporiginal}
\end{eqnarray}
where $G(p)$ is the full electron propagator, $D(q)$ is the full boson propagator, and $\Gamma_{\mathrm{v}}(p+q,p)$ is the the full vertex function. The abbreviation for integration measure is
\begin{eqnarray}
\int_{q} \equiv \int\frac{d^{3}q}{(2\pi)^3} =
T\sum_{m}\int\frac{d^{2}\mathbf{q}}{(2\pi)^2}.
\end{eqnarray}
The DS equation of the full boson propagator $D(q)$ has the form
\begin{eqnarray}
D^{-1}(q) = D_{0}^{-1}(q)-\Pi(q), \label{eq:dsefq}
\end{eqnarray}
where the polarization function is
\begin{eqnarray}
\Pi(q) = g^{2}\mathrm{Tr}\int_{p} {f}^{2} \left(\m{p}+\f{\m
q}{2} \right) \sigma_{3}G(p) \Gamma_{\mathrm{v}}(p+q,p)G(p+q),
\label{eq:dsepiq} \nonumber \\
\end{eqnarray}
where
\begin{eqnarray}
\int_{p} \equiv \int\frac{d^{3}p}{(2\pi)^3} =
T\sum_{n}\int\frac{d^{2}\mathbf{p}}{(2\pi)^2}.
\end{eqnarray}

The DS equations of $G(p)$ and $D(q)$ cannot be solved without a detailed knowledge of the vertex function $\Gamma_{\mathrm{v}}(q,p)$. Currently, it is not possible to determine the exact form of $\Gamma_{\mathrm{v}}(q,p)$. In the literature, a commonly adopted approach is to introduce the bare-vertex approximation, which amounts to setting $\Gamma_{\mathrm{v}}(q,p) \rightarrow g\sigma_{3}$, even though there is no Migdal theorem ensuring the smallness of vertex corrections. Under such a bare-vertex approximation, the DS equations of $G(p)$ and $D(q)$ become self-closed, leading to the well-known ME equations \cite{Migdal, Eliashberg}. The ME theory has achieved great success in the theoretical description of phonon mediated superconductivity in ordinary metals \cite{Allen, Carbotte, Marsiglio19, Hauckap20}, and also has been applied to study unconventional superconductivity driven by the quantum fluctuations of various order parameters \cite{Lederer15, Chubukov16, Klein18, Klein19, Wuyiming19, Damia20, Damia21, Chubukovap20, Esterlis25}. The results obtained by solving ME equations are expected to be reliable if the vertex corrections are negligible.

For electron-phonon systems, the Migdal theorem \cite{Migdal} ensures that the vertex corrections are strongly suppressed by the small factor $\lambda \omega_{D}/E_{F} \leq 1$, where $\lambda$ is a dimensionless coupling parameter and $\omega_{D}$ is Debye frequency. If the condition $\lambda \omega_{D}/E_{F} \leq 1$ is not fulfilled, vertex corrections may play a significant role and cannot be naively neglected. For instance, Esterlis \emph{et al.} found a breakdown of ME theory \cite{Esterlis18} in a two-dimensional Holstein model with $\lambda > 0.4$. In some real superconductors, the Fermi energy $E_{F}$ is quite small due to the extraordinarily low density of carriers. A notable example is SrTiO$_{3}$ \cite{Fernandes20}, where the ratio $\omega_{D}/E_{F} \gg 1$. In these superconductors, the ME formalism is believed to be invalid even though $\lambda$ is small.

The effective model for nematic quantum criticality lacks a small parameter analogous to the ratio $\omega_{D}/E_{F}$ in electron-phonon systems. Thus, the vertex corrections are only negligible when the Yukwa coupling parameter $g$ is sufficiently small. Nevertheless, in the absence of a satisfactory method of handling vertex corrections, the ME theory is still widely applied to study nematic quantum criticality. In addition to omitting vertex corrections, a couple of other approximations have been adopted to further simplify the complex ME equations.

The energy-momentum dependence of the polarization $\Pi(q)$ determines the effective strength of Yukawa coupling. Unfortunately, the concrete expression of $\Pi(q)$ can hardly be obtained. In earlier works, $\Pi(q)$ is usually computed within the perturbation theory. At the level of RPA \cite{Metlitski2010, Lederer15, Chubukov16, Klein18, Klein19, Wuyiming19, Damia20, Damia21}, the one-loop polarization is given by
\begin{eqnarray}
\Pi_{\mathrm{1}}(q) = g^{2}\mathrm{Tr}\int\frac{d^{3}k}{(2\pi)^3}f^2\left(\m{p}+\f{\m{q}}{2}\right)\sigma_{3} G_{0}(k)\sigma_{3}G_{0}(k+q),\n \\
\end{eqnarray}
which behaves as
\begin{eqnarray}
\Pi_{\mathrm{1}}(q) = -g^2{f}^{2}\left(\m{p}+\f{\m q}{2}\right)\frac{|\omega_{m}|}{|\mathbf{q}|}
\label{eq:rpapi1q}
\end{eqnarray}
in the region $|\omega_{m}|\ll |\mathbf{q}|$. This polarization has a simple form and can be used to compute the one-loop electron self-energy $\Sigma(\epsilon_{n},\mathbf{p})$ at zero temperature. At the Fermi momentum $\mathbf{p}_{\mathrm{F}}$, the electron self-energy is found to display the following frequency dependence:
\begin{eqnarray}
\Sigma_{1}(\epsilon_{n}) \propto |\epsilon_{n}|^{2/3}.
\label{eq:electronselfenergy}
\end{eqnarray}

\begin{figure}[H]
\centering
\includegraphics[width=0.95\linewidth]{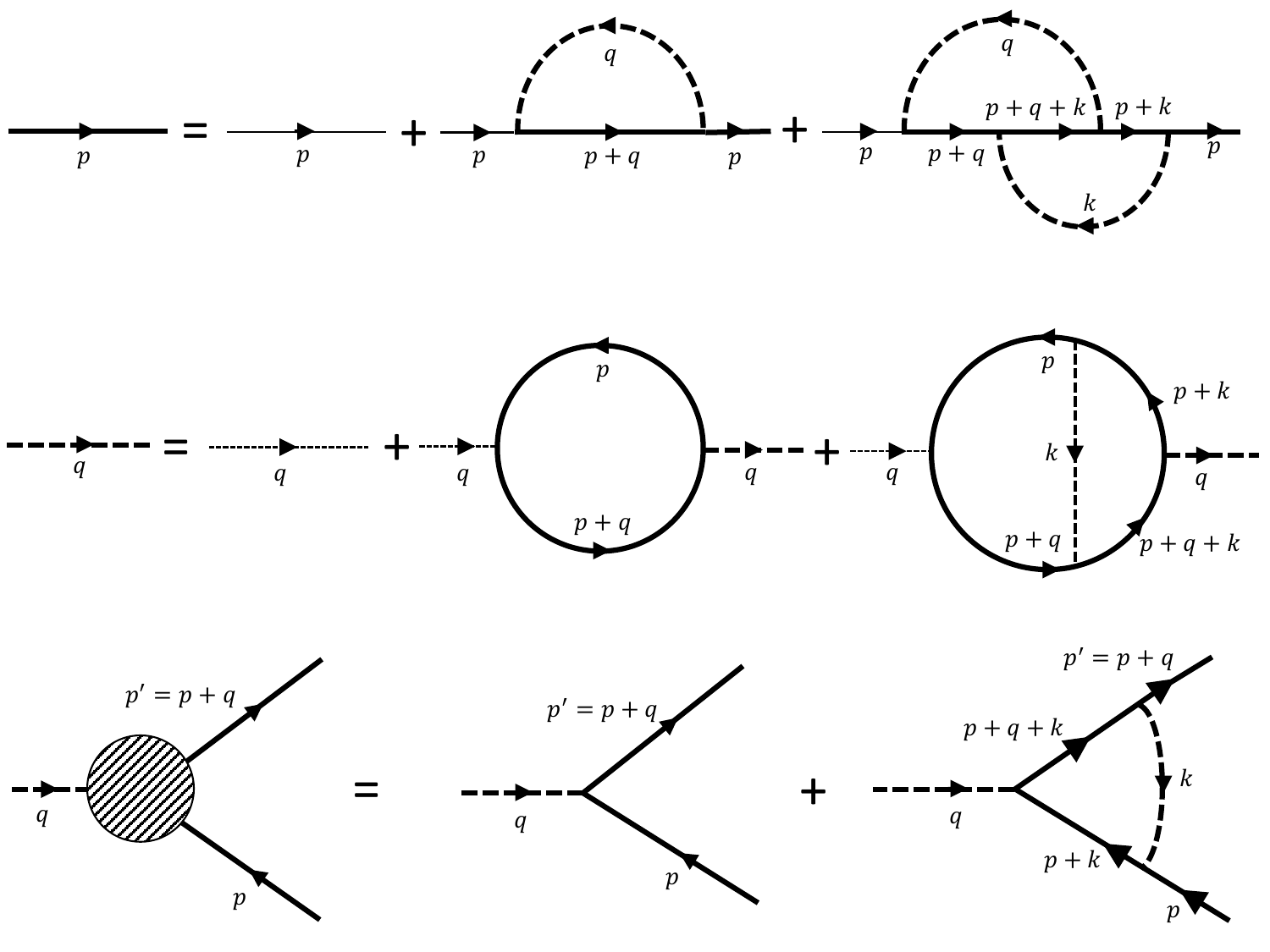}
\caption{Diagrams for the DS equations of $G(p)$, $D(q)$, and
$\Gamma_{\mathrm{v}}^{\mathrm{1L}}(p+q,p)$. Thin and thick solid
(dashed) lines represent free and renormalized electron (boson)
propagators, respectively.} \label{fig:feynmandiagram}
\end{figure}

The above expressions of $\Pi_{\mathrm{1}}(q)$ and
$\Sigma_{1}(\epsilon_{n})$ are both obtained based on several
approximations, which omit certain potentially important
contributions. For instance, the RPA polarization $\Pi_{1}(q)$
neglects the feedback effects of NFL behavior and Cooper pairing on
nematic fluctuations. The electron self-energy
$\Sigma_{1}(\epsilon_{n})$ is only valid at low frequencies and at
$T=0$, and does not account for the feedback from the pairing gap.

Another common approximation is to consider only the electrons
located exactly on Fermi surface \cite{Lederer15, Chubukov16,
Klein18, Klein19, Wuyiming19, Damia20, Damia21}, implemented by
fixing the absolute value of the electron momentum $\mathbf{p}$ at
the Fermi momentum $\mathbf{p}_{\mathrm{F}}$. Using this
approximation, the pairing function is independent of momentum,
which significantly reduces the difficulty of numerical
computations. However, while this approximation is justified for the
FL regime, it could result in considerable uncertainties in the gap
size and $T_{c}$ in the NFL regime, where the electrons occupy a
sizeable portion of states off the Fermi surface. To determine the
gap and $T_{c}$ with higher precision, it is essential to maintain
the contributions from the full range of electron momenta.

Furthermore, the nonlinear ME equations are often linearized near
$T_{c}$ based on the fact that the gap vanishes continuously as $T
\rightarrow T_{c}$. Then the determination of $T_{c}$ is converted
into the calculation of the largest eigenvalue of a linear operator.
It is clear that linearization is invalid at temperatures well below
$T_{c}$. In Refs.~\cite{Klein19, Wuyiming19, Wanghj18, Damia21}, the
nonlinear ME gap equations were solved directly, revealing new
behaviors that cannot be derived from solutions of linearized
equations. Nevertheless, these studies still adopted other
approximations mentioned above.

The purpose of the present work is to examine how the previously
obtained results on $T_{c}$ are changed when the widely used
approximations are removed. We will: (1) replace the simple
RPA-level polarization $\Pi^{\mathrm{1L}}(q)$ with the
self-consistent $\Pi(q)$ determined from the DS equation of $D(q)$;
(2) retain the mutual influence between Landau damping and Cooper
pairing so as to treat NFL behavior and superconductivity on an
equal footing; (3) incorporate the contributions from the electrons
resided not only on but also away from the Fermi surface; (4)
include the contributions of vertex corrections in the determination
of the superconducting $T_{c}$.

We now consider vertex corrections corresponding to triangle loop
diagram. These vertex corrections have been inserted into the
electron self-energy in studies on phonon-mediated superconductivity
\cite{Schrodi20, Margine25}, while the free phonon propagator
remaining unchanged. Here, in order to treat the electrons and the
nematic fluctuations equally, we will incorporate vertex corrections
into the DS equations of both $G(p)$ and $D(q)$. At the lowest
(one-loop) order, the DS equation for the triangle vertex function
$\Gamma_{\mr{v}}^{1\mathrm{L}}(p+q,p)$ has the expression
\begin{eqnarray}
\Gamma_{\mr{v}}^{1L}(p',p)&=& g\sigma_3 + \f{g^3}{f
\big(\f{\m{p+p'}}{2}\big)}\int_{k}f\big(\m{p'}+\f{\m
k}{2}\big)\nonumber \\
&&\times f\big(\f{\m{p+p'}}{2}+\m k\big)f\big(\m{p}+\f{\m
k}{2}\big)\nonumber \\
&&\times \sigma_{3}D(k)G(p'+k)\sigma_3G(p+k)\sigma_{3}.
\label{eq:trianglevertex}
\end{eqnarray}
This $\Gamma_{\mr{v}}^{1\mathrm{L}}(p+q,p)$ enters into the DS
equations of $G(p)$ and $D(q)$, which then become
\begin{eqnarray}
G^{-1}(p) &=& G_{0}^{-1}(p) +g\int_{q} {f}^{2}
\left(\m{p}+\f{\m{q}}{2}\right)D(q) \nonumber \\
&&\times \sigma_{3}G(p+q)\Gamma_{\mr{v}}^{\mathrm{1L}}(p+q,p),
\label{eq:dseofgp} \\
D^{-1}(q) &=& D_{0}^{-1}(q) - g\mathrm{Tr}\int_{p} {f}^{2}
\left(\m{p}+\f{\m q}{2} \right) \nonumber \\
&& \times \sigma_{3}G(p)
\Gamma_{\mathrm{v}}^{\mathrm{1L}}(p+q,p)G(p+q),
\label{eq:dsedq}.
\end{eqnarray}
A diagrammatic illustration of these DS equations is plotted in
Fig.~\ref{fig:feynmandiagram}. Most of the uncertainties arising
from the aforementioned approximations can be largely mitigated if
the three integral equations of $\Gamma^{\mathrm{1L}}(p+q,p)$,
$G(p)$, and $D(q)$, given by Eqs.~(\ref{eq:trianglevertex}),
(\ref{eq:dseofgp}), and (\ref{eq:dsedq}), respectively, are
numerically solved in a self-consistent manner.

The renormalized electron propagator $G(p)$ can be generically
expressed as follows
\begin{eqnarray}
G(p) = \frac{1}{A_{1}(\epsilon_{n},\mathbf{p})\epsilon_{n}\sigma_{0}
- A_{2}(\epsilon_{n},\mathbf{p})\xi_{\mathbf{p}}\sigma_{3} +
\Phi(\epsilon_{n},\mathbf{p})\sigma_{1}}. \nonumber \\
\label{eq:fullgpsimpler}
\end{eqnarray}
As demonstrated by Nambu \cite{Nambu}, the $\sigma_{2}$ term needs
not be considered. Quantum many-body effects and their interplay are
encoded in the mass renormalization function
$A_{1}(\epsilon_{n},\mathbf{p})$, the chemical potential
renormalization $A_{2}(\epsilon_{n},\mathbf{p})$, and the pairing
amplitude $\Phi(\epsilon_{n},\mathbf{p})$. In particular, NFL
behavior manifests itself in the frequency dependence of
$A_{1}(\epsilon_{n},\mathbf{p})$.

Substituting Eq.~(\ref{eq:fullgpsimpler}) into
Eq.~(\ref{eq:trianglevertex}), Eq.~(\ref{eq:dseofgp}), and
Eq.~(\ref{eq:dsedq}), we decompose the equations of $G(p)$ into
three self-consistent integral equations of
$A_{1}(\epsilon_{n},\mathbf{p})$, $A_{2}(\epsilon_{n},\mathbf{p})$,
and $\Phi(\epsilon_{n},\mathbf{p})$ given by
\begin{widetext}
\begin{eqnarray}
A_{1}(\epsilon_{n},\m p) &=& 1+\frac{T}{\epsilon_{n}}
\sum_m\int\f{d^{2}\m q}{(2\pi)^2} \tilde{\mathcal{K}}_v(\omega_m,{\mathbf{q}},
\epsilon_n,{\mathbf{p}})
A_{1}(\epsilon_{n}+\omega_{m},\m{p+q})(\epsilon_{n}+\omega_{m}),
\label{eq:3deqofa1}\\
A_2(\epsilon_{n},\m p) &=& 1-\f{T}{\xi_{\m p}}\sum_m\int \f{d^{2}\m
q}{(2\pi)^2}\tilde{\mathcal{K}}_{v}(\omega_m,{\mathbf{q}},
\epsilon_n,{\mathbf{p}})A_2(\epsilon_{n}+\omega_{m},\m{p+q})\xi_{\m{p+q}},
\label{eq:3deqofa2} \\
\Phi(\epsilon_{n},\m p) &=& T\sum_m\int \f{d^{2}\m q}{(2\pi)^{2}}
\tilde{\mathcal{K}}_{v}(\omega_m,{\mathbf{q}},
\epsilon_n,{\mathbf{p}})
\Phi(\epsilon_{n}+\omega_{m},\m{p+q}).
\label{eq:3deqofphi}
\end{eqnarray}
Here, we have defined the following kernel function $\tilde{\mathcal{K}}_{v}$:
\begin{eqnarray}
\tilde{\mathcal{K}}_{v}(\omega_m,{\mathbf{q}},\epsilon_n,{\mathbf{p}})
&=& \frac{g^2{f}^{2}\left(\m{p}+\f{\m q}{2}
\right)}{\w_{m}^{2}/v_{\mathrm{B}}^2+|\m q|^{2}+r +
\Pi(\omega_{m},\m q)} \nonumber \\
&& \times \frac{\Gamma_{\mr{v}}^{\mathrm{1L}}(\e_n+\w_m,
\m{p+q},\e_n,\m p)}{A_1^2(\epsilon_{n}+\omega_m,\mathbf{p+q})
(\epsilon_{n}+\omega_{m})^{2} + A_{2}^{2}(\epsilon_{n} +
\omega_m,\mathbf{p+q})\xi_{\m{p+q}}^2 + \Phi^{2}(\epsilon_{n} +
\omega_m,\mathbf{p+q})}.\label{eq:kernelfunction}
\end{eqnarray}
The polarization function $\Pi(\omega_{m},\m q)$ that appears in
Eq.~(\ref{eq:kernelfunction}) satisfies its own equation:
\begin{eqnarray}
&&\Pi(\omega_{m},\m q) = g^2T\sum_{n}\int \f{d^2\m p}{(2\pi)^2}
\f{{f}^{2}\left(\m{p}+\f{\m
q}{2}\right)\Gamma_{\mr{v}}^{\mathrm{1L}}(\e_n +
\w_m,\m{p+q},\e_n,\m
p)}{A_{1}^{2}(\epsilon_{n},\m{p})\epsilon_{n}^{2} +
A_{2}^{2}(\epsilon_{n},\m{p})\xi_{\m{p}}^{2} +
\Phi^2(\epsilon_{n},\m{p})}\n \\
&& \times
\frac{A_{1}(\epsilon_{n}+\omega_m,\m{p+q})(\epsilon_{n}+\omega_m)
A_{1}(\epsilon_{n},\m{p})\epsilon_{n}-A_2(\epsilon_{n}+\omega_m,\m{p+q})
\xi_{\m{p+q}}A_2(\epsilon_{n},\m{p})\xi_{\m{p}} +
\Phi(\epsilon_{n}+\omega_m, \m{p+q}) \Phi(\epsilon_{n},
\m{p})}{A_{1}^2(\epsilon_{n}+\omega_m,\m{p+q})(\epsilon_{n}+\omega_{m})^{2}
+ A_2^2(\epsilon_{n}+\omega_m,\m{p+q})\xi_{\m{p+q}}^2 +
\Phi^2(\epsilon_{n}+\omega_m,\m{p+q})}. \label{eq:3deqofpi}
\end{eqnarray}
In principle, $\Pi(\omega_{m},\m q)$ is expected to vanish in the
zero-energy and zero-momentum limits, as required by conservation
laws and the divergence of nematic propagator at the QCP where
$r=0$. However, generic theoretical analysis reveals that
$\Pi(\omega_{m}= 0,\m q \rightarrow 0)$ is not well defined, since
its value depends on the order of integration \cite{Metzner03,
Thier11, Torroba14, WangHuaJia17,Klein20, Holder14, Debbeler23,
Debbeler24, Franz02, Mross10, Grover25}. Specifically,
$\Pi(\omega_{m},\m q)$ may vanish or approach a certain constant
$\Pi_{0}$ in the limits $\omega_{m}=0$ and $\m q \rightarrow 0$.
Extensive studies have identified a constant $\Pi_{0}$ in various
contexts, including $U(1)$ gauge theories \cite{Franz02, Mross10},
NFL behavior due to nematic critical fluctuations \cite{Metzner03,
Thier11, Torroba14, WangHuaJia17,Klein20}, charge- or spin-density
wave quantum criticality \cite{Holder14, Debbeler23, Debbeler24},
and interacting fermion system with a Bose surface \cite{Grover25}.
A common approach to addressing this problem is to subtract the
constant $\Pi_{0}$ from $\Pi(\omega_{m},\m q)$, which ensures that
the effective polarization vanishes at $\omega_{m}= 0$ and $\m q
\rightarrow 0$. Here, we adopt this approach and redefine the
polarization function as
\begin{eqnarray}
\Pi(\omega_{m},\m q) - \Pi(\omega_{m}=0,\m q \rightarrow 0)
\rightarrow \Pi(\omega_{m},\m q).
\end{eqnarray}

The vertex function $\G_{\mr{v}}^{\mathrm{1L}}(\e_{n'},\m p',\e_n,\m
p)$ fulfills its own integral equation:
\begin{eqnarray}
\G_{\mr{v}}^{\mathrm{1L}}(\e_{n'},\m p',\e_n,\m p) &=& g + g^{3} T \sum_{l}\int \f{d^2\m k}{2\pi^{2}} \f{f^{-1}\big(\f{\m{p+p'}}{2}\big) f(\m{p'}+\f{\m k}{2})f(\f{\m{p+p'}}{2}+\m k)f(\m{p}+\f{\m k}{2})}{A_{1}^{2}(\epsilon_{n} + \w_l,\m{p+k})\epsilon_{n}^{2}+A_{2}^{2}(\epsilon_{n}+\w_{l},\m{p+k})\xi_{\m{p+k}}^2 +\Phi^2(\epsilon_{n}+\w_l,\m{p+k})}\n \\
&& \times \frac{\left[\w_{l}^{2}/v_{\mathrm{B}}^2+|\m k|^{2}+r+\Pi(\omega_{l},\m k)\right]^{-1}}{A_{1}^2 (\epsilon_{n'}+\w_l,\m{p'+k})(\epsilon_{n'}+\omega_{l})^{2} +A_2^2(\epsilon_{n'}+\omega_l,\m{p'+k})\xi_{\m{p'+k}}^2 +
\Phi^2(\epsilon_{n'}+\omega_l,\m{p'+k})}\n \\
&&\times \Big[A_{1}(\epsilon_{n'}+\omega_{l},\m{p'+k})(\epsilon_{n'}+\omega_{l})
A_{1}(\epsilon_{n}+\w_{l},\m{p+k})(\epsilon_{n}+\w_{l})-A_2(\epsilon_{n'}+\omega_{l},\m{p'+k})\xi_{\m{p'+k}} \n \\
&&\times A_2(\epsilon_{n}+\w_{l},\m{p+k})\xi_{\m{p+k}}+\Phi(\epsilon_{n'}+\omega_{l},\m{p'+k})\Phi(\epsilon_{n}+\w_{l},\m{p+k})\Big].
\label{eq:eqgamma1l}
\end{eqnarray}
The superconducting gap is defined as
\begin{eqnarray}
\Delta(\epsilon_{n},\mathbf{p}) = \frac{\Phi(\epsilon_{n},
\mathbf{p})}{A_{1}(\epsilon_{n},\mathbf{p})}.
\end{eqnarray}

To facilitate numerical calculations, all the involved quantities
are made dimensionless after performing the following re-scaling
transformations:
\begin{eqnarray}
&& A_{1} \rightarrow A_1, \quad A_{2} \rightarrow A_{2}, \quad T
\rightarrow \f{T}{\mr{E_F}}, \quad \e_n\rightarrow
\f{\e_n}{\mr{E_F}},\quad \w_m\rightarrow \f{\w_m}{\mr{E_F}}, \\
&& \Phi\rightarrow \f{\Phi}{\mr{E_F}}, \quad g^2\rightarrow
\f{g^2}{\mr{E_F}}, \quad \m p \rightarrow \f{\m p}{\mr{p_F}}, \quad
r\rightarrow \f{r}{\mr{p^2_F}}, \quad \xi_{\m p}\rightarrow
\f{\xi_{\m {p}}}{\mr{E_F}} = \f{\f{{\m
p}^{2}-\mr{p_{F}^{2}}}{2m}}{\f{\mr{p}_F^{2}}{2m}} = {\m p}^{2}-1.
\end{eqnarray}
\end{widetext}

\section{Numerical results \label{sec:gapsymmetry}}

The effective model defined by Eq.~\ref{eq:effectivemodel} has been
studied by Lederer \emph{et al.} \cite{Lederer15} and Chubukov
\emph{et al.} \cite{Klein18, Klein19, Wuyiming19} based on certain
approximations, including the RPA polarization $\Pi_{1}(q)$ given by
Eq.~(\ref{eq:rpapi1q}) and the Fermi-surface approximation. Lederer
\emph{et al.} \cite{Lederer15} focused on the BCS-ME regime,
characterized by a relatively large value of $r$, and found an
obvious enhancement of superconductivity due to nematicity in both
$s$-wave and $d$-wave channels. Klein and Chubukov \cite{Klein18}
considered the deep NFL regime at the nematic QCP with $r=0$. The
renormalization factor $A_{1}(p)$ adopted in Ref.~\cite{Klein18} was
derived from the lowest self-energy $\Sigma_{1}(\epsilon_{n})$
expressed in Eq.~(\ref{eq:electronselfenergy}). In
Refs.~\cite{Klein19, Wuyiming19}, the non-linear gap equation for
$\Delta(\epsilon_n)$ was solved without using the linearization
adopted in Ref.~\cite{Klein18}, although other conventional
approximations were retained. Recently, the role of vertex
corrections within an electron-nematic model has been examined by
Zhang \textit{et al.} \cite{Zhangss24} based on a simplified Ward
identity. However, a systematic investigation of the influence of
vertex corrections on the superconductivity driven by nematic
fluctuations is still lacking.

Here, we re-visit the same model by going beyond some of previously
used approximations. We will solve the non-linear integral equations
given by Eqs.~(\ref{eq:3deqofa1})-(\ref{eq:eqgamma1l}) using the
iteration method demonstrated in Ref.~\cite{Liu21}. We will not
introduce the Fermi-surface approximation, but instead allow the
electron momentum $|\mathbf{p}|$ to span the full interval
$[0,2p_{\mathrm{F}}]$. In all our calculations, the dimensionless
coupling constant is fixed at $g=0.45$.

To compute the integrals numerically, we discretize the
three-dimensional space spanned by the Matsubara frequency
$\epsilon_{n}$, the absolute value of momentum $|\mathbf{p}|$, and
the angle $\theta$ into small grid cells. These three variables are
sampled at $42$, $51$, and $164$ points, respectively. The task of
solving self-consistent nonlinear integral equations are
significantly beyond the capabilities of typical computer resources,
primarily because the vertex function
$\G_{\mr{v}}^{\mathrm{1L}}(\e_{n'},{\m p}',\e_n,{\m p})$ has six
variables. To reduce the computational burden in numerical
integration, here we adopt a selective interpolation strategy to
handle the large number of grid points. Among all the $42$ frequency
points, we retain only a reduced set of $12$ points and approximate
the other $30$ points using the linear interpolation scheme, which
effectively preserves the accuracy while greatly diminishing the
computational cost. The same strategy can be applied to deal with
$|\mathbf{p}|$ and $\theta$. Among the $51$ discrete points for
$|\mathbf{p}|$ and $164$ discrete points for $\theta$, we retain
$11$ points for $|\mathbf{p}|$ and $36$ points for $\theta$. In
particular, these points are retained adaptively, with a higher
density allocated to the regions where the vertex function exhibits
a pronounced variation. The computational burden can be further
alleviated by exploiting the inherent symmetries of the vertex
function. For instance, the integral can be evaluated only for a
minimal set of representative directions, and the results can be
mapped to all symmetry-equivalent directions via appropriate
operations. Through all these optimizations, the total computational
time has been reduced by four to five orders of magnitude, bringing
it down to an acceptable value.

To estimate the error induced by linear interpolation, we have
computed $T_{c}$ within ME theory for $g=0.45$ and $r= 10^{-2}$. The
interpolated and fully resolved values are $T_{c} = 0.0333$ and
$T_{c}=0.0335$, respectively. Therefore, linear interpolation leads
to an error of $\approx 0.6 \%$. For the same parameters $g=0.45$
and $r= 10^{-2}$, the value of $T_{c}$ obtained with vertex
corrections is $T_{c}=0.0285$, which is $\approx 15 \%$ lower than
the bare-vertex result $T_{c}=0.0335$. Since the interpolation error
is two orders of magnitude smaller than the vertex-correction
effect, we adopt linear interpolation as a well-justified
approximation.

While the self-consistent integral equations are equally applicable
for any value of $r$, the numerical computations become increasingly
more difficult as $r$ decreases. For smaller values of $r$, it is
necessary to choose more points for each integration variable. The
computational time required to reach convergence is dramatically
increased when $r$ is lowered. To ensure the precision of the
results, the value of $r$ is restricted to the range of $r\ge
10^{-3}$.

After solving the integral equations
(\ref{eq:3deqofa1})-(\ref{eq:eqgamma1l}), we have obtained the full
frequency, momentum, and angular dependence of all four functions of
$A_{1}(p)$, $A_{2}(p)$, $\Phi(p)$, and $\Pi(q)$. Their
three-dimensional visualizations are rather cumbersome and do not
convey clear information. Therefore, we will not show these color
maps and instead focus on the symmetry of superconducting gap and
the behavior of $T_{c}$, which are experimentally accessible and can
be compared with existing data.

\subsection{Superconducting gap symmetry \label{sec:symmetry}}

Although the form factor $\hat{f}(\theta)$ of the Yukawa coupling
exhibits a $d$-wave angular dependence, the resulting pairing gap
does not necessarily display $d$-wave symmetry \cite{Lederer15}. If
the ME equations are linearized in the limit $T \rightarrow T_{c}$,
which amounts to dropping $\Phi$ from the denominators, the value of
$T_{c}$ for each pairing channel is determined by the largest
eigenvalue of the resultant linear equation. Klein and Chubukov
\cite{Klein18} have found that $s$-wave and $d$-wave gaps are nearly
degenerate at nematic QCP, but the $T_{c}$ for $s$-wave symmetry is
slightly higher than that for $d$-wave symmetry. Recently, a highly
anisotropic $s$-wave superconducting gap has been observed
\cite{dsneto24} near the nematic QCP of FeSe$_{1-x}$S$_{x}$, which
is qualitatively consistent with the gap symmetry predicted in
Ref.~\cite{Klein18}.

To determine the pairing symmetry, here we solve the nonlinear
integral equations by setting various initial values of $\Phi(p)$
and analyze the angular dependence of the convergent solutions for
the superconducting gap $\Delta(p)$. One could assign a
$\theta$-independent constant initial value or an arbitrary
$\theta$-dependent function to $\Phi(p)$. Under the bare vertex
approximation, our computations show that the nonlinear equations
have only two possible solutions: an extended $s$-wave gap that is
positive for all angles and a sign-changing $d_{x^{2}-y^{2}}$-wave
gap. A $d_{x^{2}-y^{2}}$-wave solution is obtained if and only if a
pure $d_{x^{2}-y^{2}}$-wave initial value of $\Phi(p)$ is assumed.
However, this pure $d_{x^{2}-y^{2}}$-wave solution is fragile. When
the initial value of $\Phi(p)$ is different from a pure
$d_{x^{2}-y^{2}}$-wave form, the solution of $\Phi(p)$ inevitably
evolves into an extended $s$-wave gap. Moreover, once vertex
corrections are incorporated, the extended $s$-wave pairing emerges
as the sole stable solution for all possible initial values of
$\Phi(p)$.

\begin{widetext}

\begin{figure}[H]
\centering
\includegraphics[width=0.9\textwidth]{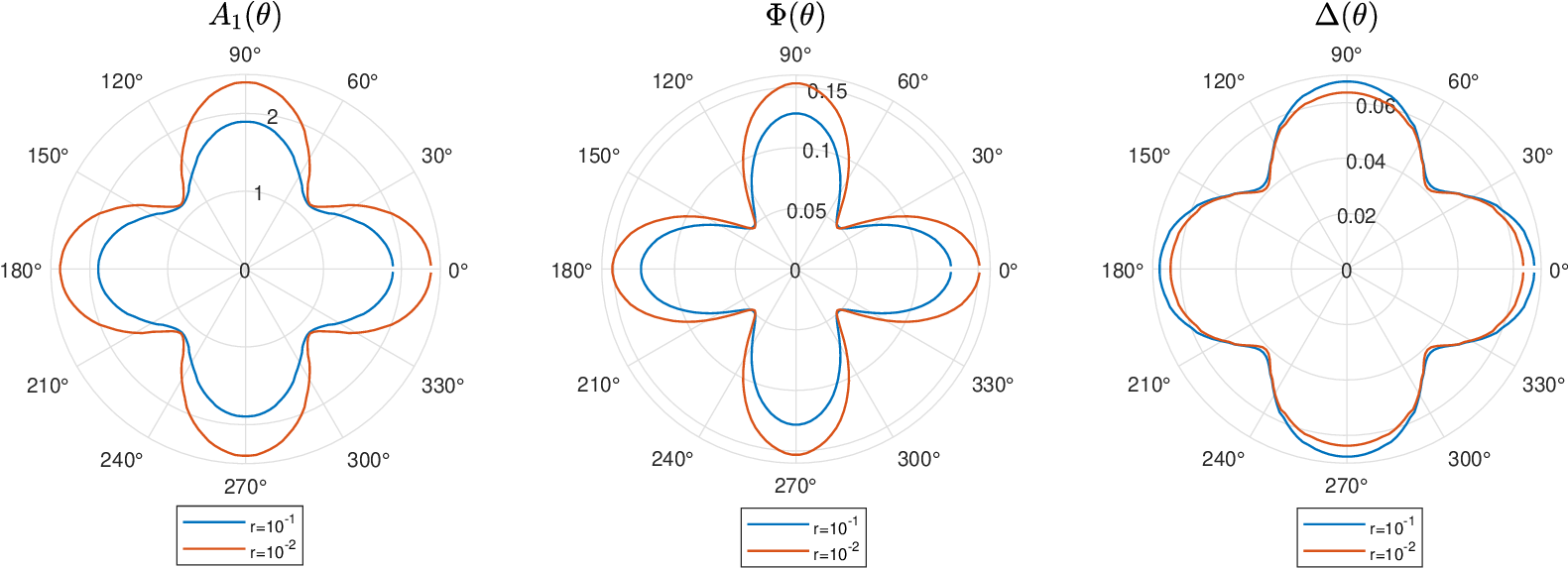}
\caption{Angular dependence of $A_1(\theta)$, $\Phi(\theta)$, and
$\Delta(\theta)$ at the frequency $\epsilon_{n}=\pi T$ and the Fermi
momentum $\mathbf{p}_{\mathrm{F}}$. Other model parameters are
$g=0.45$ and $T=0.01$. Blue and orange-red curves correspond to $r =
10^{-1}$ and $r = 10^{-2}$, respectively.}
\label{fig:thetadependence}
\end{figure}

\end{widetext}

In Fig.~\ref{fig:thetadependence}, we depict the detailed
$\theta$-dependence of $A_{1}(\theta)$, $\Phi(\theta)$, and
$\Delta(\theta)$ at the frequency $\epsilon_{n}=\pi T$ and at the
Fermi momentum. The $s$-wave gap obtained in our calculations aligns
with the results of Klein and Chubukov \cite{Klein18} and is also
consistent with the observation reported by Nag \emph{et al.}
\cite{dsneto24}. The gap observed by Nag \emph{et al.}
\cite{dsneto24} is highly anisotropic and nearly nodal. In
comparison, the gap shown in Fig.~\ref{fig:thetadependence} exhibits
merely a moderate anisotropy, probably because the values of $r$
used in our calculations are not sufficient small.

Recent muon spin relaxation measurements \cite{Matsuura23} have
provided experimental evidence that, for $x \leq 0.22$, the
FeSe$_{1-x}$S$_x$ superconductor hosts a superconducting state in
which time-reversal symmetry is spontaneously broken. Our numerical
computations yield no indication of such an unconventional pairing
state. Probably, the simple single-band model under consideration
lacks certain degrees of freedom or interactions that are essential
for realizing time-reversal-symmetry breaking.

\subsection{Dependence of $T_{c}$ on $r$ \label{sec:gaponr}}

To clarify how the critical temperature $T_{c}$ is influenced by
various theoretical ingredients, we compare the results obtained
under three different approximations.

$\bullet$ VERTEX: $T_{c}$ is determined by solving the fully
self-consistent DS equations given by
(\ref{eq:3deqofa1})-(\ref{eq:eqgamma1l}) without any additional
approximations.

$\bullet$ ME: The DS equations are simplified to the ME equations
for $G(p)$ and $D(q)$ after adopting the bare vertex approximation,
i.e. replacing the vertex function
$\G_{\mr{v}}^{\mathrm{1L}}(\e_{n'},\m p',\e_n,\m p)$ with
$g_{0}\sigma_{3}$.

$\bullet$ MEFS: The ME equations are further simplified by fixing
the electron momentum at the Fermi momentum (i.e., Fermi-surface
approximation).

\begin{figure}[H]
\centering
\includegraphics[width=0.44\textwidth]{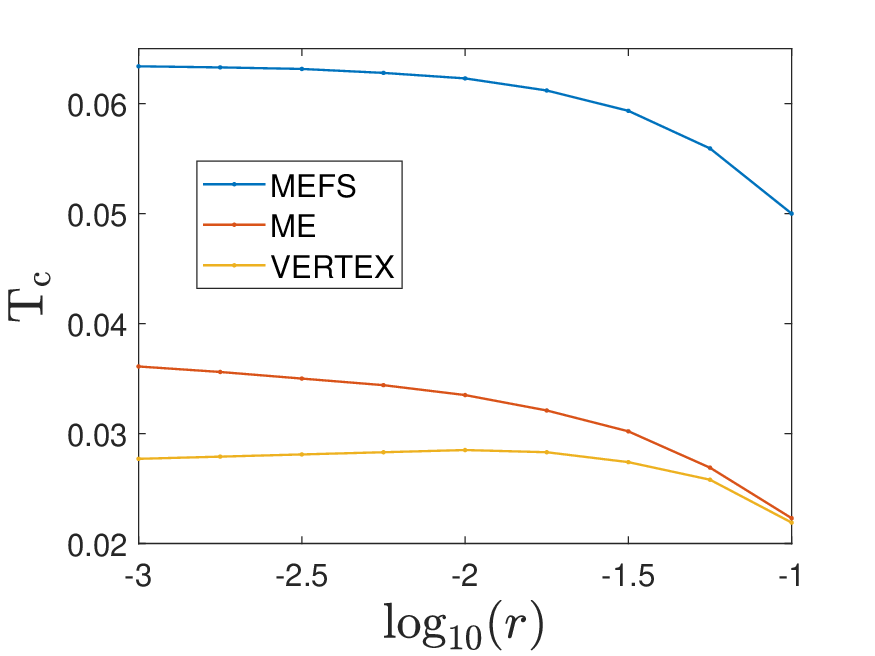}
\caption{Superconducting transition temperature $T_{c}$ as a
function of the parameter $r$ for $g=0.45$. The blue, orange-red,
and yellow curves correspond to the MEFS, ME, and VERTEX
approximations, respectively. $T_{c}=0$ if $r$ becomes sufficiently
large.} \label{fig:tconr}
\end{figure}

The resulting $T_{c}$ for each case is shown in
Fig.~\ref{fig:tconr}. It can be seen that neglecting vertex
corrections tends to overestimate $T_{c}$. The introduction of Fermi
surface approximation leads to a more significant overestimation of
$T_{c}$. An apparent indication is that the contributions of the
electrons off the Fermi surface make important contributions to the
formation of superconductivity.

Under the bare vertex approximation, for either ME or MEFS curve,
the corresponding $T_{c}$ is a monotonously decreasing function of
$r$. This implies that superconductivity is always enhanced as the
system gets closer to the nematic QCP. However, the behavior of
$T_{c}(r)$ is qualitatively changed when the vertex corrections are
taken into account. According to the VERTEX curve shown in
Fig.~\ref{fig:tconr}, as $r$ decreases from $10^{-1}$ to $10^{-3}$,
$T_{c}$ first increases but then decreases after passing through its
maximal value. Therefore, the function $T_{c}(r)$ determined in the
presence of vertex corrections exhibits a non-monotonic dependence
on the distance to the nematic QCP.

\begin{figure}[H]
\centering
\includegraphics[width=0.4\textwidth]{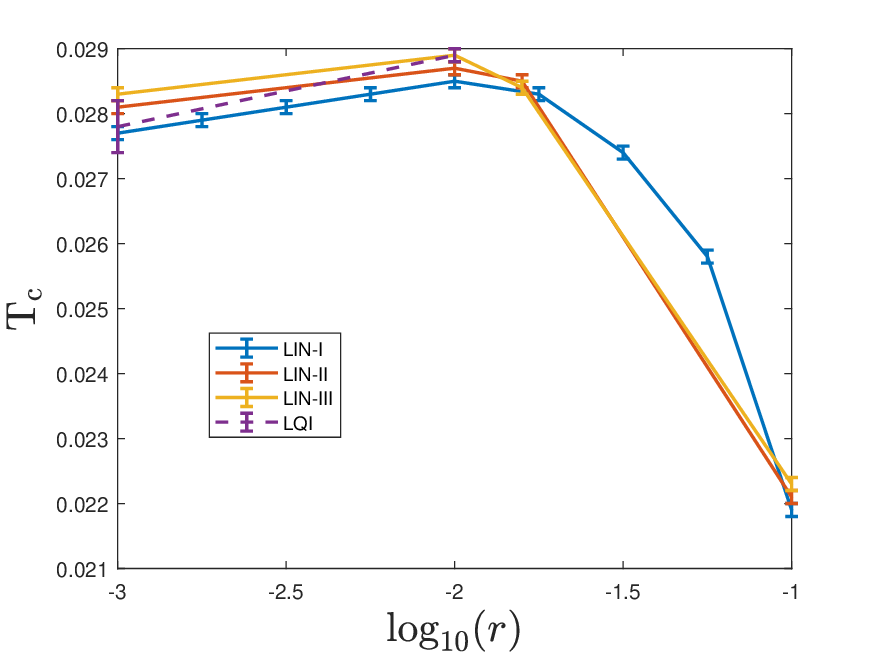}
\caption{Curves of $T_{c}(r)$ with error bars. Here, $g=0.45$.}
\label{fig:tcerrorbars}
\end{figure}

Since the value of $T_{c}$ is only moderately suppressed as $r$ is
decreased in the range $r < 10^{-2}$, it is necessary to examine
whether the non-monotonic behavior of $T_{c}(r)$ is changed if other
interpolation methods are adopted. We employ three distinct linear
interpolation methods, labeled as LIN-I, LIN-II, and LIN-III, to
compute $T_{c}$, and list the original results in Table
\ref{table1}. The total numbers of sampling points for
$(\epsilon_{n},|\mathbf{p}|,\theta)$ are fixed at $(42,51,164)$ for
each method. However, the number of selected sparse sampling points
are different. The corresponding $T_{c}(r)$ curves are presented in
Fig.~\ref{fig:tcerrorbars}. Although different interpolation schemes
result in slight variations in the values of $T_{c}$, the
non-monotonic behavior of $T_{c}(r)$ is observed for all
interpolation schemes. Furthermore, we have also gone beyond the
linear interpolation and employed a more accurate local quadratic
interpolation (LQI) \cite{Davis1963}. The computational time
required for LQI is roughly ten times longer than that for linear
interpolation methods. Since our goal is to verify the reliability
of the non-monotonicity of $T_{c}(r)$, we have chosen only two
representative values of $r$ in LQI calculations: $r=10^{-2}$ and
$r=10^{-3}$. The results presented in Table \ref{table1} and
Fig.~\ref{fig:tcerrorbars} demonstrate that the non-monotonicity of
$T_{c}(r)$ is a robust feature rather than an artifact of the linear
interpolation approximation.

Owing to computational limitations, we have not achieved converged
solutions of the nonlinear integral equations for $r < 10^{-3}$. The
behavior of $T_{c}(r)$ in this regime remains unknown. To get
reliable results of $T_{c}$ in this region, the space spanned by
$\epsilon_{n}$, $|\mathbf{p}|$, and $\theta$ should be discretized
into smaller grids, which will substantially increase the
computational time.

\begin{figure}[H]
\centering
\includegraphics[width=0.42\textwidth]{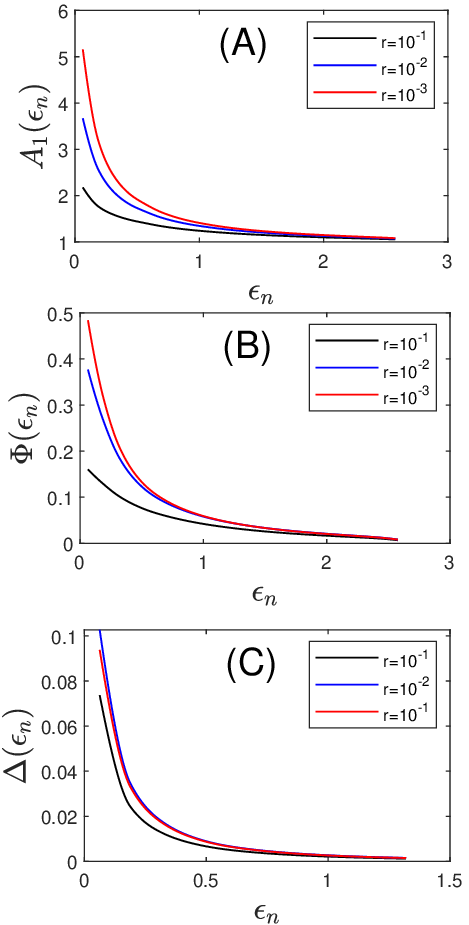}
\caption{The frequency dependence of $A_1(\epsilon_{n})$,
$\Phi(\epsilon_{n})$, and $\Delta(\epsilon_{n}) =
\Phi(\epsilon_{n})/A_1(\epsilon_{n})$ in the $\theta=0$ direction
for electrons restricted on the Fermi surface. The results are
obtained for $g=0.45$ and $T=0.02$.}
\label{fig:A1PhiDelta}
\end{figure}

It is important to find out the origin of the striking non-monotonic
evolution of $T_{c}(r)$. We fix the electron momenta on the Fermi
surface, i.e., $\mathbf{p} = \mathbf{p}_{\mathrm{F}}$, and compare
the $r$-dependence of $A_{1}(\epsilon_{n})$ with that of
$\Phi(\epsilon_{n})$. From the results shown in
Fig.~\ref{fig:A1PhiDelta}~(a)-(b), we find that the reduction of $r$
gives rise to a monotonic increase in both $A_{1}(\epsilon_{n})$ and
$\Phi(\epsilon_{n})$. Thus, the NFL behavior and Cooper pairing are
both enhanced as the nematic QCP is approached. Nevertheless, it is
important to notice that the superconducting gap
$\Delta(\epsilon_{n})$ is the ratio
$\Phi(\epsilon_{n})/A_{1}(\epsilon_{n})$ rather than
$\Phi(\epsilon_{n})$. The gap size can be suppressed when
$A_{1}(\epsilon_{n})$ is enhanced more significantly than
$\Phi(\epsilon_{n})$. To display the $r$-dependence of gap size, we
also plot the function of $\Delta(\epsilon_{n})$ for $r = 10^{-1}$,
$r = 10^{-2}$, and $r = 10^{-3}$ in Fig.~\ref{fig:A1PhiDelta}(c).
One can see that the superconducting gap determined at $r=10^{-2}$
is larger than those determined at $r=10^{-1}$ and $r=10^{-3}$. The
non-monotonic $r$-dependence of the gap size eventually results in
the non-monotonic $r$-dependence of $T_{c}$. These results suggest
the presence of an intricate interplay between NFL behavior and
Cooper pairing near nematic QCP.

\begin{table}[H]
    \centering
    \setlength{\tabcolsep}{6pt}
    \caption{Results of $T_{c}$ determined using four different interpolation methods. The numbers given in the columns of $\epsilon_{n}$, $|\mathbf{p}|$, and $\theta$ represent the numbers of the selected sparse sampling points. The error in the LQI results for $r=0.001$ becomes larger because the required computational effort increases substantially.}
    \begin{tabular}{c c c c c c c}
    \toprule[1pt]
    Method & $T_c$  & Error &$\epsilon_n$  &$|\mathbf{p}|$& $\theta$ & $r$ \\
    \midrule[0.8pt]
    LIN-I & 0.0219 & ±0.0001 &12 & 11 &36 & $10^{-1}$ \\
    LIN-II & 0.0221 & ±0.0001 &12 & 11 &20 & $10^{-1}$ \\
    LIN-III & 0.0223 & ±0.0001 &10 & 11 &36 & $10^{-1}$ \\
    \midrule[0.6pt]
    LIN-I & 0.0283 & ±0.0001 &12 & 11 &36 & $2\times 10^{-2}$ \\
    LIN-II & 0.0285 & ±0.0001 &12 & 11 &20 & $2\times 10^{-2}$ \\
    LIN-III & 0.0284 & ±0.0001 &10 & 11 &36 & $2\times 10^{-2}$ \\
    \midrule[0.6pt]
    LIN-I & 0.0285 & ±0.0001 &12 & 11 &36 & $10^{-2}$ \\
    LIN-II & 0.0287 & ±0.0001 &12 & 11 &20 & $10^{-2}$ \\
    LIN-III & 0.0289 & ±0.0001 &10 & 11 &36 & $10^{-2}$ \\
    \midrule[0.6pt]
    LIN-I & 0.0277 & ±0.0001 &12 & 11 &36 & $10^{-3}$ \\
    LIN-II & 0.0281 & ±0.0001 &12 & 11 &20 & $10^{-3}$ \\
    LIN-III & 0.0283 & ±0.0001 &10 & 11 &36 & $10^{-3}$ \\
    \midrule[0.8pt]
    LQI & 0.0289 & ±0.0001 &12 & 11 &36 & $10^{-2}$ \\
    LQI & 0.0278 & ±0.0004 &12 & 11 &36 & $10^{-3}$ \\
    \bottomrule[1pt]
    \end{tabular}\label{table1}
\end{table}

\subsection{Comparison with experiments}

Early theoretical works \cite{Yamase13, Scalapino15, Lederer15,
Metlitski15} predicted a pronounced enhancement of the
superconducting $T_{c}$ at the nematic QCP, where the pairing
interaction mediated by nematic fluctuations is strongest. This
prediction has been extensively tested in iron-based
superconductors. In the BaFeAs family, the nematic order usually
coexists with antiferromagnetic spin-density-wave (SDW) order, and
the resulting entanglement of the nematic and magnetic fluctuations
obscures their individual effects on superconductivity. It appears
more suitable to explore the relationship between nematicity and
superconductivity in FeSe superconductors due to the absence of SDW
order at ambient pressure.

Over the past decade, there has been a surge of experimental studies
\cite{Coldea17, Bohmer17, Hosoi16, Sunjp16, Matsuura17, Sato18,
Hanaguri18, Wiecki18, Coldea20, Gallais21, Mukasa21, Shibauchi22,
Occhialini23, Mukasa23, Matsuura23, Coldea24, Dingqp25, Shibauchi25}
on the properties of superconductivity emerging around the nematic
QCP in FeSe$_{1-x}$S$_x$ and FeSe$_{1-x}$Te$_x$ systems. Contrary to
theoretical expectations, the value of $T_{c}$ is typically not
obviously boosted at the nematic QCP, irrespective of whether
nematic order is destroyed by chemical substitution or the variation
of pressure.

Some recent works have attributed the absence of significant
enhancement of $T_{c}$ at nematic QCP to the effects of
nematoelastic coupling \cite{Paul17, Labat17, Gallais21,
Shibauchi22, Freire22, Fernandes25}. Labat and Paul \cite{Labat17}
defined a dimensionless parameter $r_{0}$ to represent the effective
strength of nematoelastic coupling and then demonstrated that
$T_{c}$ is strongly enhanced only when the condition $r_{0} \ll
(U/V)^{2}$ is fulfilled, where $U$ measures the pairing interaction
strength generated by nematic fluctuations and $V$ collects all the
non-nematic contributions (such as magnetic fluctuations or
phonons). When this inequality is reversed, namely $r_{0} \gg
(U/V)^{2}$, the increase of $T_{c}$ is modest even at the QCP.
Applying this criterion to FeSe$_{1-x}$S$_x$, where strong magnetic
fluctuations yield a sizeable $V$, the absence of appreciable
$T_{c}$ enhancement at the nematic QCP $x=0.17$ can be naturally
explained \cite{Gallais21, Shibauchi22}. The situation is distinct
in FeSe$_{1-x}$Te$_x$, which displays a clear superconducting dome
\cite{Mukasa21, Shibauchi22, Mukasa23} in the vicinity of nematic
QCP defined at $x_{c}=0.50$ at ambient pressure. Since no magnetic
fluctuations were detected within the same doping region
\cite{Mukasa21, Shibauchi22}, implying that $V=0$, it is likely that
the superconducting dome is solely induced by nematic fluctuations.
According to Labat-Paul criterion \cite{Labat17, Shibauchi22},
$T_{c}$ should be strongly enhanced at the QCP, regardless of
whether the nematoelastic coupling is weak or strong.

However, the superconducting dome observed in FeSe$_{1-x}$Te$_x$
exhibits an anomalous shape \cite{Mukasa21, Shibauchi22, Mukasa23}:
$T_{c}$ is not maximal at the nematic QCP ($x_{c} = 0.50$), but
peaks at a higher Te concentration $x_{p}\approx 0.60$. This
displacement of the optimal doping ($x_{p}\approx 0.60$) away from
the nematic QCP, together with the unusual suppression of $T_{c}$ at
the QCP, is incompatible with the theories predicting the maximum
$T_{c}$ to occur precisely at the QCP \cite{Yamase13, Scalapino15,
Lederer15, Metlitski15}. Furthermore, this observation cannot be
reconciled with the scenario of nematoelastic coupling
\cite{Labat17}, which allows only for an increase in $T_{c}$ at the
QCP. In Ref.~\cite{Mukasa21}, the authors speculated that such a
difference may originate from the more significant enhancement of
quasiparticle damping effects than the pairing interaction.

According to our results, the transition temperature $T_{c}$
determined after including vertex corrections displays a
non-monotonic $r$-dependence: $T_{c}$ is peaked at a finite $r_{m}$
rather than at the nematic QCP ($r=0$). In the region of
$0<r<r_{m}$, $T_{c}$ is gradually suppressed as $r$ is lowered. This
behavior provides a possible interpretation of the salient features
$T_{c}$ observed on the disordered side of the nematic QCP in
FeSe$_{1-x}$Te$_x$, including the discrepancy between the optimal
doping and the nematic QCP and the anomalous suppression of $T_{c}$
as the system is approaching to QCP.

As illustrated in Fig.~\ref{fig:A1PhiDelta}, the decrease of $r$
enhances both the NFL behavior, embodied in $A_{1}(\epsilon_{n})$,
and the pairing strength, measured by $\Phi(\epsilon_{n})$, but the
NFL behavior is enhanced more significantly as $r\rightarrow 0$. It
turns out that the non-monotonic $T_{c}(r)$ curve is generated by
the subtle interplay of NFL decoherence and Cooper pairing of
strongly damped electrons. This lends a quantitative support to the
scenario proposed in Ref.~\cite{Mukasa21} and highlights the pivotal
role of vertex corrections in capturing the interplay of different
quantum many-body effects.

Our present analysis is applicable solely to the disordered
(tetragonal) side of nematic QCP. The ordered (orthorhombic) side
may exhibit qualitatively different properties. For instance, the
Fermi surface distortion caused by nematic order could lead to
effects absent in the disordered side. Moreover, there are
experimental evidences \cite{Hanaguri18, Sato18} suggesting that the
superconducting gap in FeSe$_{1-x}$S$_x$ has distinct structures on
the disordered and ordered sides of nematic QCP. In
FeSe$_{1-x}$Te$_x$, the superconducting dome emerging in the doping
region $x>0.30$ also displays different features from the smaller
superconducting dome observed deep inside the nematic phase
\cite{Shibauchi22}. The physical origin of such distinctions remains
elusive and awaits further research. The possible competition
between nematicity and superconductivity \cite{Hirschfeld20} further
complicates the situation.

\section{Summary and discussion \label{sec:summary}}

In summary, we studied the properties of the superconductivity
formed in the vicinity of a two-dimensional Ising-type nematic QCP.
Focusing on the disordered side of the transition, we treat the NFL
behavior and Cooper pairing on an equal footing by solving the
self-consistent nonlinear DS equations for the electron and nematic
propagators and the lowest-order vertex correction without adopting
further approximations. The only stable solution is an extended
$s$-wave gap. Our calculations showed that vertex corrections
qualitatively alter how the superconducting $T_{c}$ evolves as the
system is tuned to approach the nematic QCP from the disordered
side. Under the bare-vertex approximation, $T_{c}$ is always
enhanced. After including vertex corrections, $T_{c}$ exhibits a
non-monotonic dependence on the distance to nematic QCP and its
maximum is displaced from the QCP. this result provide a possible
explanation for the anomalous evolution of $T_{c}$ observed in the
tetragonal phase of FeSe$_{1-x}$Te$_x$. We demonstrated that the
non-monotonic evolution of $T_{c}$ stems from a delicate interplay
between NFL decoherence and Cooper pairing. Such an interplay is
hidden under the bare-vertex approximation and can be unveiled only
when the vertex corrections are taken into account, suggesting the
necessity of incorporating vertex corrections into the ME theory.

The analysis presented in this paper can be improved along the
following directions:

(i) Our vertex corrections to the Yukawa coupling are presently
limited to the lowest-order triangle diagrams. While even this
minimal inclusion already reveals qualitatively new features in
$T_{c}$, a systematic treatment of higher-order contributions is
required. It is especially important to develop efficient schemes
that can incorporate vertex corrections while preserving the
Ward-Takahashi identities.

(ii) Nematic quantum criticality is currently treated based on a
single-band model. This simplification is ideal for examining how
vertex corrections modify the ME theory. However, it cannot capture
the complicated multi-orbital electronic structures found in
realistic materials, such as FeSe$_{1-x}$S$_x$ and
FeSe$_{1-x}$Te$_x$. To provide a faithful account of these
superconductors, the single-band model should be extended to
incorporate the multi-orbital degrees of freedom \cite{Kang18,
Chubukov25} and the competition and coexistence of nematic and
superconducting orders \cite{Hirschfeld20}. It would be of interest
to study whether including the multi-orbital effects can lead to
time-reversal symmetry broken superconducting states.

(iii) Solving the nonlinear integral equations with higher precision
demands a more efficient algorithm. Very close to the nematic QCP
($r < 10^{-3}$), both NFL behavior and Cooper pairing are governed
by small-$|\mathbf{q}|$ scattering, which requires a much finer grid
in the $(\epsilon_{n},|\mathbf{p}|,\theta)$ space. However, as the
mesh is refined, the computational time required to reach convergent
results of $T_{c}$ is substantially increased. Furthermore, the
nonlinear equations would become even more complicated after
considering the multi-orbital effects. Balancing the accuracy and
the computational cost calls for a further algorithmic optimization.

\section{Acknowledgement}

We thank Hao-Fu Zhu for helpful discussions. This work is supported
by the Anhui Natural Science Foundation under Grant 2208085MA11.
J.R.W. acknowledges the support by the National Natural Science
Foundation of China under Grant 12274414. The numerical calculations
in this paper have been done on the supercomputing system in the
Supercomputing Center of University of Science and Technology of
China.

\end{sloppypar}
\end{document}